\documentclass[aps,prl,superscriptaddress,showpacs,twocolumn,floatfix]{revtex4-1}
\usepackage{graphicx}
\usepackage[usenames,dvipsnames]{color}
\usepackage{amsmath}
\usepackage{lineno}
\usepackage{url}
\usepackage{graphicx}
\newcommand{\PR}{{ Phys. Rev. }}
\newcommand{\PRL}{{ Phys. Rev. Lett. }}

\newcommand{\etal}{{\em et al.}}
\newcommand\Cornell{Cornell University, Ithaca, NY 14853 }
\newcommand\CWM{College of William and Mary, Williamsburg, VA}
\newcommand\NCCU{North Carolina Central University, Durham, NC 27707}
\newcommand\JLab{Thomas Jefferson National Accelerator Facility, Newport News, VA 23606}
\begin{document}
\preprint{APS/123-QED}
\title{A limit on the anisotropy of the one-way maximum attainable speed of the electron}
\author{W.~Bergan} \affiliation{\Cornell}
\author{M.J.~Forster} \affiliation{\Cornell}
\author{V.~Khachatryan} \affiliation{\Cornell}
\author{N.~Rider} \affiliation{\Cornell}
\author{D.L.~Rubin} \affiliation{\Cornell}
\author{B.~Vlahovic} \affiliation{\NCCU}
\author{B.~Wojtsekhowski}
\thanks{Corresponding author: bogdanw@jlab.org} 
\affiliation{\CWM} \affiliation{\JLab}
\date{\today}
\begin{abstract}
We report here the first experimental result for the anisotropy of the one-way maximum attainable speed of the electron, 
$\vec{\Delta c_{1,e}}$, obtained via the study of a sidereal time dependence of a difference between the electron and 
positron beam momenta in the CESR storage ring at Cornell University. 
At 95 percent confidence, an upper limit for the component of $\Delta \vec {c}_{1,e}/c$ perpendicular to Earth's 
rotational axis is found to be $5.5 \times 10^{-15}$.
\end{abstract}
\maketitle

\paragraph{Introduction.$-$}
The Theory of Special Relativity (TSR)~\cite{AE1905} was formulated from a few postulates whose 
experimental tests were important for the universal acceptance of this foundation of modern physics.
The principle of relativity has been confirmed by perfect agreement between the TSR predictions and experiments.
The universal value of the speed of light in inertial reference systems has been also tested experimentally.
Since 1905 the precision of the tests has improved by many orders, and now such experiments represent 
an important approach to the search for physics which has some degree of Lorentz invariance violation (LIV).
According to the original theory by A.~Einstein, the speed of light in a vacuum is isotropic
and the maximum attainable speed for all particles is equal to the speed of light.
However, it is important to differentiate between the two-way speed (average over a closed path, $c_2$),
whose isotropy for light was addressed for the first time in the Michelson-Morley experiment~\cite{MM1887}, 
and the one-way speed, $c_1$, whose isotropy is a more general postulate.

A number of later formulations of TSR are based on the postulate of relativity and
the isotropy of the two-way speed, see for review~\cite{RA1998}.
At the same time, high precision measurement of $c_1$ isotropy is a well developed approach 
for tests of LIV in several frameworks including the Standard-Model Extension (SME)~{\cite{DC1997}}.
The maximum attainable speed of elementary particles could differ from the speed of light.
The current limits on deviations of the maximum attainable speed of particles
from the speed of light and related bounds on the anisotropies of the maximum attainable speed (AMAS) 
are discussed in { Ref.~\cite{KO2011}}.

The current upper bound for the photon AMAS, $\Delta c_{2,photon}/c$ is $10^{-18}$, see Ref.~\cite{NA2015},
which is already in the domain of { Quantum Gravity (QG) effects}~\cite{DM2005}.
The more challenging for experiment, the $\Delta c_{1,photon}/c$ bound is $\sim 1.6 \times 10^{-14}$, according to Ref.~\cite{BO2010}.
A number of TSR predictions have also been tested with increasing accuracy, see Ref.~\cite{CW2014}.
For example, the test of the relativistic Doppler transformation performed with high speed atoms
provided an LIV test on the level of $10^{-9}$~\cite{BB2014}.

\paragraph{Momentum anisotropy.$-$}
In the TSR the momentum of the particle is a four-vector whose transformation between the inertial 
reference frames follows Lorentz's transformations.
The space component of the momentum is $\vec p = m \cdot \vec v / \sqrt{1 - (v/c)^2}$,
where $m$ is the particle mass, $\vec v$ is the particle velocity,
$v$ is the absolute value of $\vec v$; and $c$ is the speed of light.
When taking into account the difference between the maximum attainable speed of the particle, $c_{part}$ 
(with a possible directional anisotropy AMAS) and the speed of light, $c$, we should avoid
non-physical results for $p$, and 
the expression above should be written as: $\vec p = m \cdot \vec v / \sqrt{1 - (v/c_{1, part})^2}$,
where ${c}_{1, part}$ is the maximum attainable speed of the particle in the direction of the particle velocity $\vec v$.

The immediate prediction of the TSR is that in process which preserves 
the absolute value of particle speed, the absolute value of particle momentum is unchanged.
By testing such a prediction for different orientations of the momentum one can obtain a bound on LIV.
This method (momentum anisotropy) is especially sensitive for a particle with a large Lorentz factor,
$\gamma = 1/\sqrt{1-(v/c_{1, part})^2}$, in the expression above for the momentum,
because of a $\gamma^2$ enhancement: 
\vskip -.2 in
	\begin{equation}
\Delta p /p \,\approx\, -\gamma^2 (\Delta c_{1, part}/c),
		\label{eq:amom}
	\end{equation}	
\vskip -.05 in

where $\Delta p$ is a variation of the momentum. Similar kinematical enhancement of the sensitivity to LIV for other high $\gamma$ factor processes 
was obtained previously, see e.g. Refs.~\cite{VG1996, SC1999, BA2005}.

In the current report we present the first experiment carried out by using the momentum anisotropy 
method~\cite{BW2014}  and the obtained limit for the electron AMAS.

Independent of the high precision test of TSR, the search for sidereal time variation of the maximum 
attainable one-way speed of particles provides an interesting way to study the directional isotropy of the universe.
Tests of directional isotropy have been conducted by high precision NMR since the 1960's~\cite{HU1960}.
For a review and perspectives on { the current MR-based}  search methods, see Ref.~\cite{SA2017}.

A simple model for the $c_1$ anisotropy is based on the concept of a local ether, 
as distinct from the global ether ruled out in the famous experiment~\cite{MM1887}. 
This possible local ether moves relative to the solar system and has a tiny refractive index~\cite{DB2017}.
For example, this local ether may be related to the Cosmic Microwave Background radiation.

\paragraph{Magnetic deflection for the AMAS search.$-$} 
We implemented the momentum anisotropy method for a search of the electron AMAS.
The particle momentum was precisely measured via deflection in a transverse magnetic field.
The magnetic field transverse to the direction of motion also allows us to change the direction of the particle's 
momentum and repeat the momentum measurement, for example, when the particle is moving in the reverse 
direction (where the impact of AMAS reaches its maximum).
In other words, the 180$^\circ$ magnetic arc acts as a heavy mirror which reflects an electron (positron) elastically.

In the search for an LIV effect the conventional form of the Lorentz force needs to be corrected because its textbook 
form suppresses possible LIV contributions, for example the rotational non-invariant options.
The nature and exact form of such a correction are unknown, so in the current analysis of kinematics of the particle motion in the transverse magnetic field we are using a minimum number of parameters and considerations.
These are a vector of the particle speed $\vec{v}$ and another vector of the magnetic field $\vec{B}$, which is an axial vector.

The first assumption is a proportionality of the acceleration and the magnitude of each of these vectors - the absence
of non-linear terms.
We assume here the parity conserving nature of the acceleration of a charged particle moving 
in a transverse magnetic field, so the acceleration of the particle should be directed along the vector product $\vec{v}\times\vec{B}$.
For such a direction of the acceleration, the absolute value of the particle speed remains constant, which
is an important consideration used later in the interpretation of the experimental observable: 
the limit on a variation of the difference between the particle speed and the maximum attainable speed in 
direction of the particle motion, $v-c_{1,e}$.
We searched for a variation of the particle momentum allowed according to the QG dispersion relations~\cite{DM2005} and assumed energy conservation.

\paragraph{Experimental considerations.$-$} 

Our experiment was performed with bunched beams of high energy electrons and positrons circulating in a storage ring.
The precision of the experiment for $\Delta c_{1,e}/c$ benefits from a large value of the beam Lorentz factor, $\gamma$,
and the high precision of the measurement of the beam centroid location in the transverse direction 
(especially at high beam current in a storage ring).

The stable geometry and magnetic field of the accelerator magnets are of paramount importance in our method.
These requirements are significantly relaxed when the experiment uses two counter-propagating beams of 
a particle and an antiparticle in one set of magnets.
The difference between the counter-rotating particle momenta is relatively insensitive to drifts in the storage ring 
magnetic fields and geometry, while the sensitivity to $\Delta c_{1,e}/c$ is doubled, { assuming that the anisotropy
is the same for the electron and positron (validity of CPT for AMAS)}.
{ The equality of the masses and opposite sign and equal value of the electrical charges for the electron and positron, 
which are also important for our analysis, are confirmed by the experiments~\cite{PDG}
to a much higher precision than is essential to our experiment.}

The beam momentum, $p$, and accelerator lattice dispersion function, $\eta(s)$, where $s$ is a coordinate 
along the reference orbit of the beam, relate to the deviation of the horizontal beam position, $x(s)$, 
and to the nominal beam positions in the accelerator at location $s$ as: $x(s) - x_{nom}(s) \,=\, \eta(s) \times (p-p_{nom})/p$,
where $x_{nom}(s)$ is the horizontal closed orbit at momentum $p_{nom}$ and 
$(p-p_{nom})$ is a deviation of the momentum from its nominal value{~\cite{ES20XX}}.
Variation in the measured position difference of the two beams at $s$ thus corresponds to 
the measurement of variation in their momenta difference. 
By measurement of the beam position in an area free from accelerating elements, the beam momenta 
variations in different directions of motion can be found.
The uncertainty in the absolute position of the beams is irrelevant as we are attempting to measure 
only the time dependence of the momenta at the frequency of a sidereal rotation.
While the particle energy varies with the coordinate $s$ due to synchrotron radiation energy loss 
and beam acceleration in the RF cavities, the energy at any $s$ is stable and calculable with high precision.

The primary experimental observable is a momentum difference at time moment $t$ between the two 
counter-rotating beams defined as  $\Delta p_\pm(s,t) \,=\, \left[ x_+(t) \,-\, x_-(t) \right] /\eta(s)$, which allows us 
to search for a potential signal using the following equation (for the dipole form of AMAS):
\vskip -.2 in
	\begin{equation*}
		\Delta p_\pm (s, t) \,=\, a_\perp \times [ \, \cos(\Omega \cdot t - \Phi)  * \cos ({ \Psi(s)}) \,+
	\end{equation*}	
\vskip -.35 in
		\begin{equation}
			+ \sin(\theta_{_{CESR}}) * \sin(\Omega \cdot t - \Phi) * \sin ({\Psi(s)})],
		\label{eq:amas}
	\end{equation}
%
where $a_\perp \,=\, 2\gamma^2 \times \Delta c_{1,e}^\perp/c$ is a fit parameter,
$\gamma$ is the beam Lorentz factor, $\Delta c_{1,e}^\perp/c$ is an AMAS value,
$\Omega$ is a sidereal  frequency which is close to the Earth's rotational frequency, 
$\Phi$ is the phase which is defined by the direction of anisotropy,
${\Psi(s)}$ is the phase of the beam position monitor location at position $s$ in the storage ring, and
$\theta_{_{CESR}}$ is the geographic latitude of the Cornell Electron Storage Ring.

\begin{figure}[hb]
\vskip -.2 in
	\centering
	\includegraphics[width=0.25\textwidth, angle = 0]{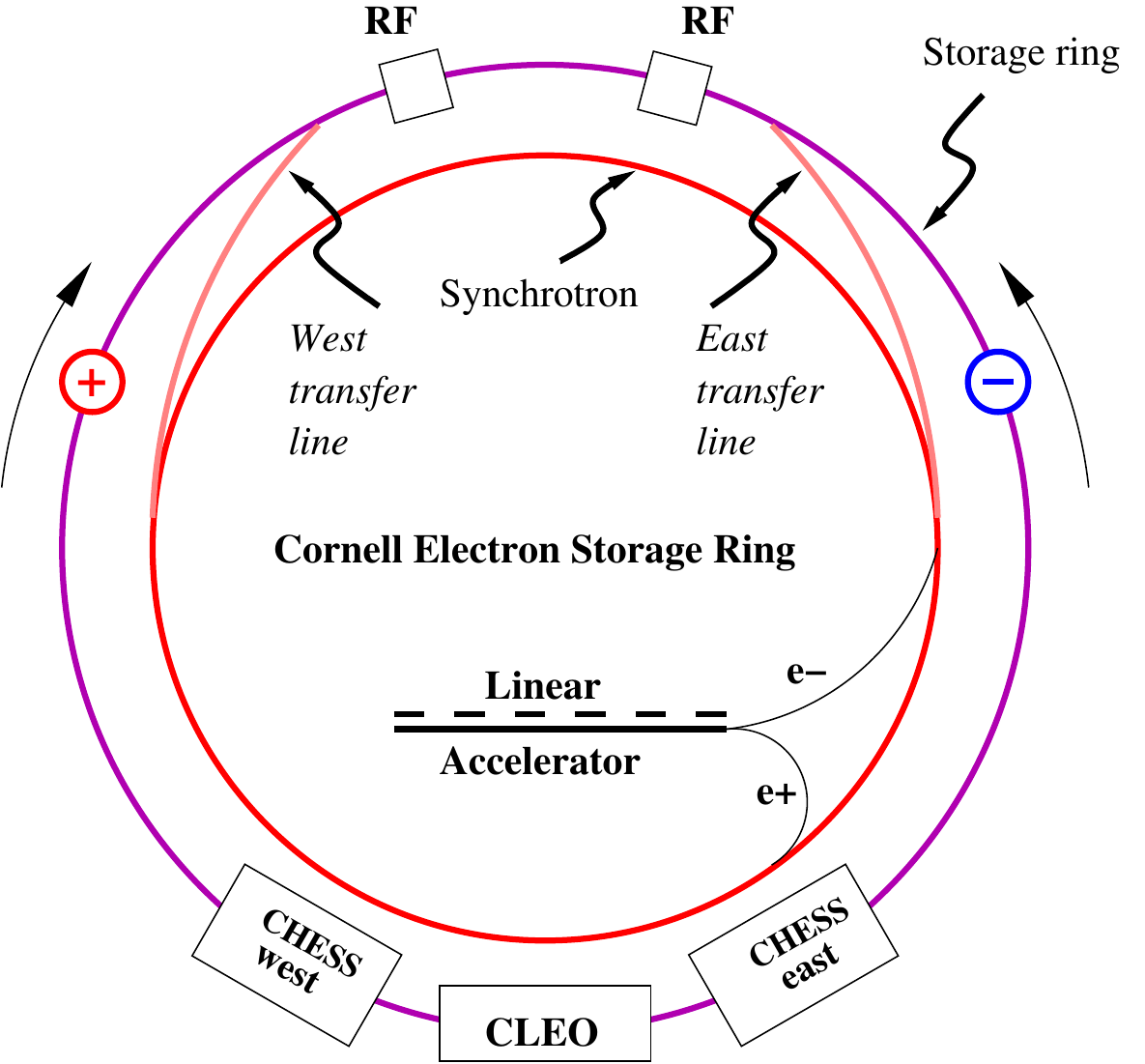} 
	\caption{CESR ring geometry and features. 
	RF indicates locations of the RF cavities. 
	CLEO indicates the location used for the collider detector and 
	CHESS stands for the Cornell High Energy Synchrotron Source.}
	\label{fig:CESR}
	\end{figure}

\paragraph{The experiment.$-$}
Our experiment used beams of electrons and positrons with an energy of 5.29~GeV 
in the Cornell Electron Storage Ring (CESR)~\cite{CESR}, Fig.~\ref{fig:CESR}. 
	
The ring layout is mirror symmetric about its north-south diameter and has a 768~m circumference. 
Its lattice is of the focusing-defocusing (FODO) type, with each half cell consisting of a dipole, quadrupole, 
and sextupole magnet.
The  design of the magnetic lattice minimizes beta functions at the collision points,
in order to mitigate current dependence of the beam-beam interaction on the closed orbits.
The impact of the remaining beam-beam interaction was investigated in the collected orbit data
and found to be much smaller than can be detected at the achieved accuracy of 
the Beam Position Monitor (BPM) and consistent with theoretical expectations.
The dispersion function has a value of between one and two meters in most of the orbit.
The RF accelerating cavities are located symmetrically near the south straight section.

The beams' electrostatic separation system and automatic beam orbit feedback system were turned off.
The ring was filled with one electron bunch and one positron bunch with $1.2 \times 10^{10}$~particles 
in each bunch corresponding to 0.75~mA. 
The collision points were chosen to be at the location of the interaction point for the former CLEO detector 
and its diametric counterpart in the north straight section.
The beam life time was an average of 5~hours.
A typical data taking cycle/fill lasted about 1-2~hours, during which the beam currents dropped to a level of 0.5-0.6~mA.

The measurement of the beam positions was performed by means of the CESR BPM system,
which includes 99 BPMs~\cite{MP2010}.
	\begin{figure}[!h]
		\vskip -.4 in
	\centering
	\includegraphics[width=0.25\textwidth, angle = 0]{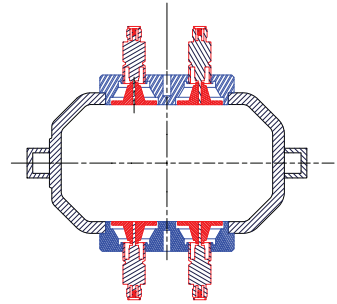}
	\vskip -.4 in
	\caption{CESR four-electrode BPM, see~\cite{DR2010}.}
	\label{fig:CESR-BPM}
	\vskip -.2 in
	\end{figure}
The CESR four-electrode BPM configuration shown in Fig.~\ref{fig:CESR-BPM} allows determination 
of both the horizontal and vertical coordinates.
The BPM signal is amplified and digitized at a sampling rate of 125~MHz. 
Each of the four electrodes has a programmable delay to ensure that the bunch signal is sampled at the peak. 
The delay step is 10~ps.
Standard DAQ allows accumulation of data for many thousands of turns from each BPM for each beam (electron or positron).
The sign of the bipolar pulse depends on the particle species.  
The signal is sampled at the peak of the leading pulse  which is of the opposite sign for electrons and positrons~\cite{MB2017}.

Two groups of measurements were performed.
In the first group (A), the data for the electron beam and the data for the positron beam were recorded 
at close but different moments in time.
These moments are shifted by 20-30 seconds due to the time needed for loading 
different versions of the local readout code into the BPM electronics.
We called the combined information from the time-consecutive position measurements for the two beams a data ``shot".
Measurements for both beams used the same set of electronics, which allowed us to reduce the impact of the 
electronic instability in the beam position difference.
The raw data included the amplitudes from four electrodes from each of 99 BPMs and the combined (electron plus positron) beam current.
Data for one shot were collected for each beam over 4000 turns in CESR, which in total covered about a 10~ms time period.
Data analyzed in this paper were obtained during several multi-hour periods in December 2016 and October 2017.
In total, in group A, 1714 data shots were taken using 35 fills of the ring.

In the second group (B), beam data were recorded  at each BPM for each sequential bunch passage 
on each turn (the time difference between readouts is less than two microseconds)
but with different electronics (amplifier and digitizer) for the two species. 
The electron beam data for 4000 turns and similar positron beam data constitute a pair of synchronized measurements (a shot).
The interval between the sequential shots was about 200 seconds.
Information for group B was collected in January 2018.
In total,  in group B, 228 data shots were taken over a 14-hour period using five fills of the ring.

\paragraph{Experiment sensitivity to the electron AMAS.$-$}
The experiment's sensitivity to the signal of interest is defined by the coordinate resolution of the BPMs, 
the lattice functions of the storage ring,  statistics of measurements,  duration of data taking and systematics due
to remaining beam-beam interaction and electronics instabilities.
The combined sensitivity of the data and effect of the analysis procedure on the detected parameters 
of the signal ($a_\perp$ and $\Phi$) was investigated by adding to the actual BPM data a ``test wave'' 
which has the shape of a potential signal.
After the adding of the test wave, a full analysis procedure was performed and the wave parameters were reconstructed.
The sensitivity studies performed for the signal with an amplitude in the range down to $a_\perp = 0.25$~ppm
show that the reconstructed amplitude is reduced by a factor of 0.74, which is a typical effect for a multi-parameter fit analysis.
The phase, $\Phi$, which defines the preferable direction for AMAS
for the lowest amplitude of the test wave was reconstructed with an accuracy of 0.5~radian.

The drift of the BPM electronics and magnets introduces systematical effects which can create artificial signals.
In the analysis of the data, as presented below, we excluded some BPMs with exceedingly large noise or drift.
In addition, for run group A, the instability of the horizontal steering magnets (kickers) introduced large orbit distortions. 
The effect of the kickers has been corrected by a fitting procedure.
The data were analyzed with various groupings which allowed us to evaluate the systematics from
the spread of the results and obtain a best estimate for an upper limit on the AMAS value.

\paragraph{Analysis of the data.$-$}

For analysis, the data from group A were arranged in five sets, each from six to eight hours long, 
and the data from group B were used in one set.
Analysis was performed independently for each of these six sets.

The analysis procedure started from an evaluation of the noise in the raw amplitudes of the four-electrode BPMs.
For this purpose we checked the correlation between the signal from an individual electrode of each BPM, 
$A_i$, of a BPM and the beam current, $I_{beam}$.
The rms of the correlation $\alpha = A_i/I_{beam}$ was analyzed. 
It was found that the relative value of the rms, $\sigma_\alpha/\alpha$, averaged over all usable BPMs, was 0.002.
When the $\sigma_\alpha/\alpha$ exceeded a four times larger value (0.008), which corresponds to about a 200~$\mu$m 
position change or about a 100~ppm momentum change, the corresponding BPM was removed from further analysis.
The applied cut eliminated on average a few BPMs of the 99 available in CESR.

For all accepted BPMs the beam horizontal position $x_+$ for positrons ($x_-$ for electrons) 
was calculated from the corresponding four amplitudes in the BPM electrodes
by using an updated iterative procedure which starts from a linearized expression~\cite{DR2010}.
The value of $x_\pm = x_+ - x_-$ provides the beam position difference.
It has contributions from i) the difference in the beam energies, which varies along the orbit but
is time independent, ii) the off-set, which is due to electronic calibration and
for most BPMs is sufficiently stable over selected periods of several hours, 
iii) the small random differences in the magnetic system between the moments 
of measurements of the coordinates of the electron and positron beams (essential for group A), 
and iv) a potential AMAS signal, which has time dependence according to the
sidereal period of the Earth's rotation and smooth variation along the beam orbit.

Within one data shot in the group A set, the time delay between the $x_+$ and $x_-$ measurements 
(20-30 seconds) sometimes leads to an orbit position change as large as 30~$\mu$m in some locations.
The most common reason is magnets with a changed value of the field (kickers).
They can be identified by analysis of the beam position as a function of $s$.
The parameters of the kicker were obtained from the fit of the data with the closed orbit function
{~\cite{ES20XX}}:
\vskip -.25 in
\begin{equation}
	f(s,j) = \theta_j \cdot \frac {\sqrt{ \beta(s) \beta(s_j)}} {2\sin(\pi Q_x)} \cos [|\phi(s) -\phi(s_j)| - \pi Q_x], 
\label{eq:fit}
\end{equation}
where $f(s,j)$ is the closed orbit function for a kicker $j$,  
$\theta_j$ is the beam deflection angle of the beam by the kicker $j$, 
$\beta(s)$ and $\beta(s_j)$ are the ring lattice beta functions 
at location $s$ and at the kicker location $s_j$, 
$Q_x$ is the horizontal betatron frequency equal to 10.55 for CESR, and
$\phi(s)$ and $\phi(s_j)$ are the lattice betatron phase advances at location $s$ and at the location of the kicker $s_j$.
The change in beam momentum due to the application of a kick of the observed amplitude in a region of finite dispersion is negligibly small.

For determination of the kickers' locations and preliminary amplitudes in a given shot, we divided 
the data set into several 30-minute time intervals (short relative to the Earth's rotation period) during 
which the stability of all other contributions to $\Delta x_\pm$ (except the kickers') is much higher than the kicker impact.
In 30 minutes the data in the group A set had about fifty shots, $N_{sh}$, which 
provide $N_{_{BPM}}*N_{sh}$ coordinate values $\Delta x_\pm$ for use in the fit.
Here $N_{_{BPM}}$ is the total number of used BPMs, which was 80 or more.
The fit per Eq.~\ref{eq:fit} was used for determination of the parameters of the kickers
and $\Delta x_\pm^{ref}$ for all other contributions combined.
The total number of fit parameters for eight kickers in each data shot and the off-sets for $N_{_{BPM}}$ BPMs is at 
a maximum of $8*N_{sh}*2 + N_{_{BPM}}$, which is small compared to the number of fit points. 

\begin{figure}[b]
	\centering
	\includegraphics[trim = 50mm 40mm 50mm 40mm, width=0.49\textwidth, angle = 0]{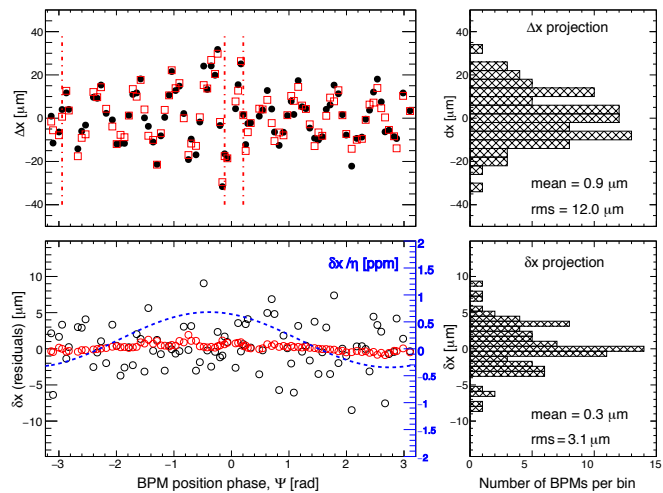} 
	\caption{Run group A kicker analysis. An example of position difference $\Delta x_\pm - \Delta x_\pm^{ref}$ vs. 
	BPM location for a typical data shot. Here the ${{\Psi}} = 2\pi \cdot s/P$, where $P$ is the perimeter of the orbit.
	The upper left panel shows the raw data as black points and the fit values as open red squares with three kickers 
	whose intensities and locations (indicated by vertical red dashed lines) were obtained from the fit of 
	a closed orbit function. 
	The upper right plot shows the distribution of the $\Delta x_\pm$ values and its rms.
	The lower panel shows the distribution of the residual $\delta x_\pm$ (after subtraction of the kickers'
         contribution) as open black circles, the signal fit function ($\delta x/\eta$) as a dashed blue curve 
         (see the scale on the right),  the signal contribution to  $\delta x_\pm$ as open red circles. 
         The corresponding projection of open black circles shown on the right side has an rms of 3.1~$\mu$m.}
	\label{fig:ddx_run1}
		\vskip -.25 in
	\end{figure}

Figure~\ref{fig:ddx_run1} shows $\Delta x_\pm$ vs. the BPM location along the orbit 
in the storage ring for a typical data shot, residual $\delta x_\pm$,
and the corresponding rms.
The significant variations of the $\Delta x_\pm$, shown in the upper panel of Fig.~\ref{fig:ddx_run1},  are due to the kickers.
The rms for $\Delta x_\pm$ is about 12~$\mu$m.
The fit of the closed orbit function (Eq.~\ref{eq:fit}) with several kickers
(the actual number of kickers was defined by the significance of the $\chi^2$ improvement) 
allowed us to find the residual $\delta x_\pm = \Delta x_\pm  - \sum f(s,j)$, where $\sum f(s,j)$ 
is the sum of the closed orbit functions for all observed kickers. 
The $\delta x_\pm$ values have a typical rms of 3~$\mu$m as is shown for one data shot
in the lower panel of Fig.~\ref{fig:ddx_run1}.

After the kickers were identified, their number and locations in each shot were fixed, and the fit was redone 
with optimization of the amplitudes of the kickers for each shot, off-sets, $\Delta x_\pm^{ref}$, for each BPM
(fixed in each group of 30-minute time intervals), the AMAS amplitude and phase 
(fixed within a given several-hour measurement set).

For group B the analysis above was not needed because the synchronized readout of BPMs 
completely suppressed the kicker effect. 
At the same time group B required additional electronics channels which added noise.
Overall the data from groups A and B have similar accuracies.

The second step of the analysis is the same for groups A and B and includes the following:
The data were fitted by a two-parameter function of the signal and one parameter per BPM $\Delta x_\pm^{ref}$ as:
\vskip -.2 in
\begin{equation*}
	\Delta x_\pm^{*} = \eta^{-1}(s) \times a_\perp \times [ \cos(\Omega \cdot t - \Phi)  * \cos({\Psi}(s)) \,+\, 
	\end{equation*}
\vskip -.30 in
	\begin{equation}
+ \, \sin(\theta_{_{CESR}}) * \sin(\Omega \cdot t - \Phi) * \sin ({\Psi}(s))] \,+\, \Delta x_\pm^{ref}.
\label{eq:fit_final}
\end{equation}
\vskip -.05 in
For all data sets in the analysis of the AMAS signal we calculated the realtime phase $\Phi$ 
relative to midnight, October 22, 2017, at the geographic location of CESR.

\begin{figure}[!]
	\centering
	\hskip -.0 in
	\includegraphics[trim =5mm 0mm 0mm 5mm, width=0.51\textwidth]{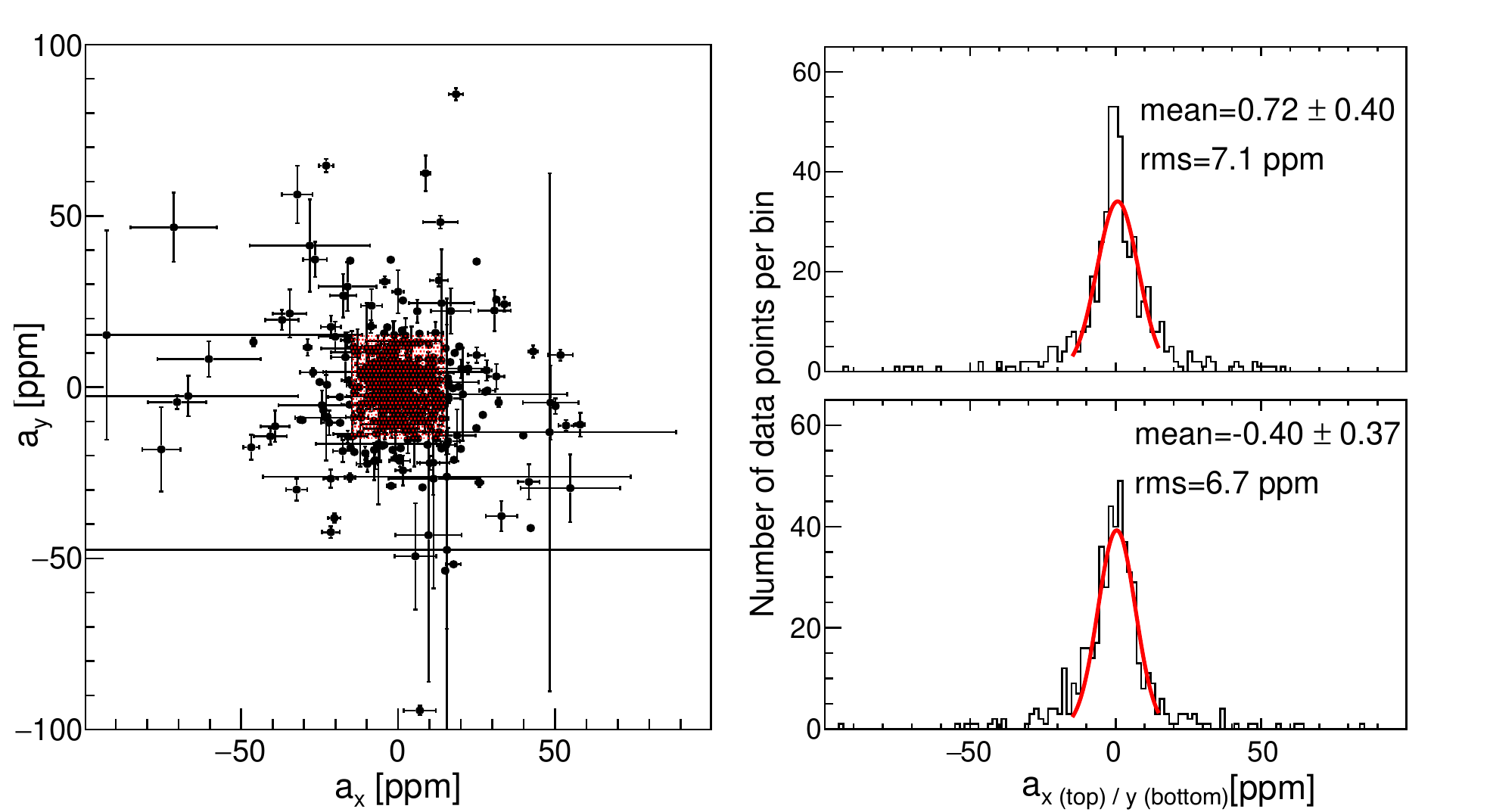} 
	\vskip -.0 in
	\caption{One-BPM fit results for the AMAS function parameters.
	The pink contour 30~ppm $\times$ 30~ppm on the left plot 
	contained 359 data points out of the 500 obtained.
	The distributions in the right side plots were fitted by a Gaussian function for the central areas of $\pm 15$ ppm.}  
\label{fig:one-BPM_fit}
\end{figure}

In the fit of the AMAS signal we took into account that some of the BPMs have a significant drift.
We determined the rate of BPM drifts by doing one-BPM fits per Eq.~\ref{eq:fit_final},  for
which the results are plotted in Fig.~\ref{fig:one-BPM_fit}.
For presentation of the search results below we used the $a_{\perp,x}$ and $a_{\perp,y}$ which 
are the components of a projection of the anisotropy vector $\vec a$ to the plane of the CESR ring 
along the geographical meridian and orthogonal to it, respectively.

The mean values for the one-BPM analysis are $a_{\perp,x} = 0.72$~ppm and 
$a_{\perp,y} = -0.40$~ppm with the rms of 7~ppm.
It is easy to see that the tails in the distribution of the widths are dominated by the systematics. 
The systematic uncertainty is much larger than the statistical uncertainty for the individual data points.
Data points (corresponding BPMs) with a extra large deviation (above 15~ppm) were excluded from the next step of the analysis. 
The accuracy of the average values (0.72/0.40~ppm) is close to expected for these 500 data points. 

\begin{figure}[ht]
	\centering
		\vskip -.15 in
	\includegraphics[trim = 0mm 10mm 10mm 0mm, width=0.3\textwidth, angle = 0]{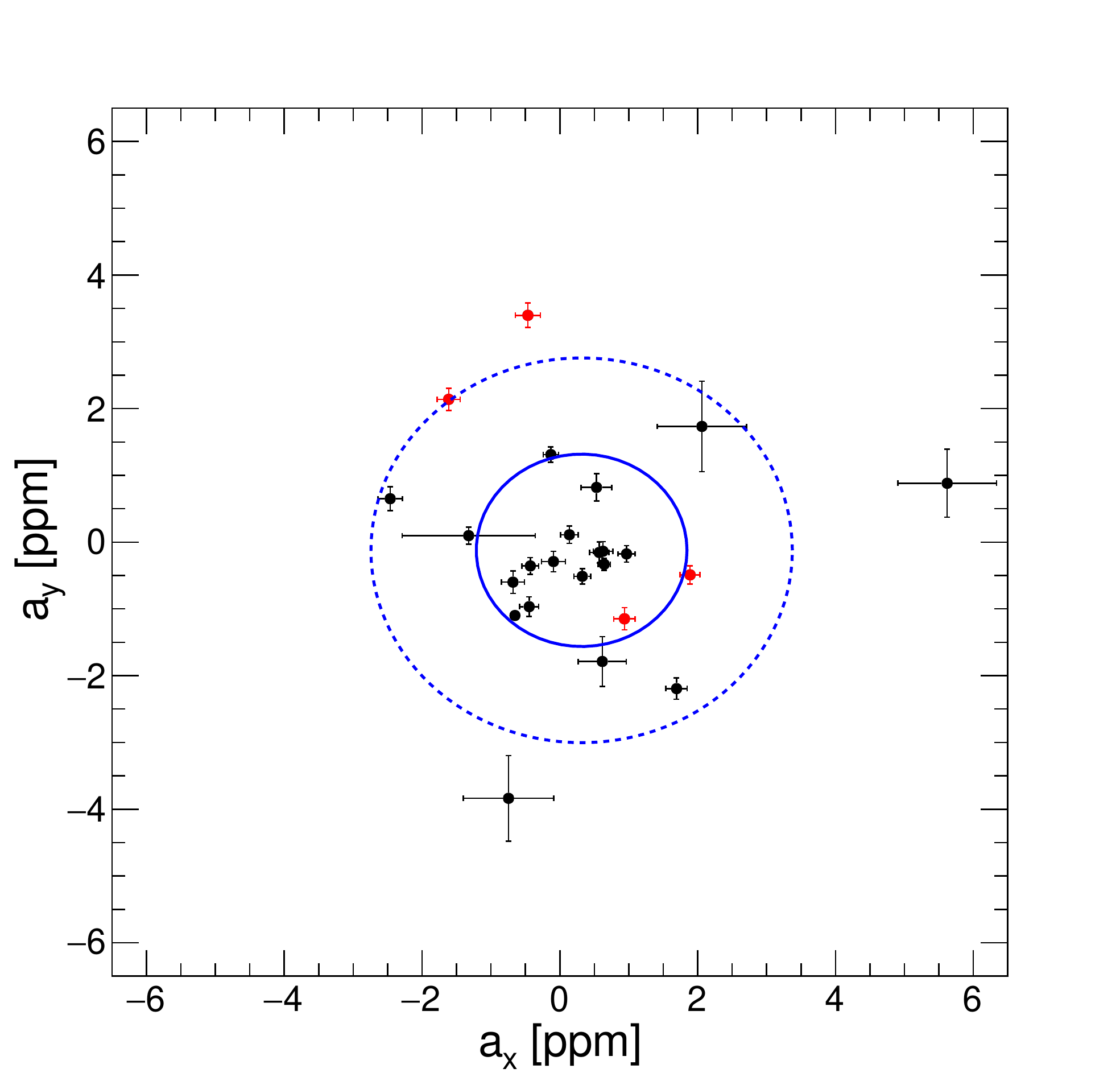} 
\caption{The results for all six measurements presented by 24 data points. 
The points in black and red show the results from the run groups A and B, respectively. 
The error bars at individual points represent statistical accuracy. The RMS is calculated from the distribution of the data points. 
The mean values are $a_{\perp,x} = 0.32 \pm 0.31$~ppm and  $a_{\perp,y} = -0.12 \pm 0.29$~ppm. 
The solid and dashed blue contours show the borders of one and two sigma exclusion areas, respectively, per one data point out of the 24 obtained.}
\label{fig:result}
\end{figure}

The final method for combining of the data is based on four groups of BPMs.
In group \#1 we used  every fourth BPM among those selected for the analysis starting from \#1,
then took \#5, \#9 and so on. In group \#2 the starting BPM is \#2, then \#6, \#10  and so on.
Each group includes about 20 BPMs distributed almost uniformly around the ring.
In each of six sets of measurements we fitted the signal by using those four groups of BPMs.
A total of 24 data points were obtained and the results are shown in Fig.~\ref{fig:result}.
The combined fit of these data points leads to the values
$a_{\perp,x} = 0.32 \pm 0.31$~ppm and  $a_{\perp,y} = -0.12 \pm 0.29$~ppm,
where $a_{\perp,x} = a_{\perp} \times \cos(\Phi)$ and 
$a_{\perp,y} = a_{\perp} \times \sin(\Phi)$ per Eq.~\ref{eq:amas},
which corresponds to the value of a combined fit amplitude $a_\perp = 0.34 \pm 0.42$~ppm
and phase $\Phi = -0.36$~radian.
\paragraph{Summary.$-$}

The search for the anisotropy of the one-way maximum attainable speed of the electron was performed 
by means of precision monitoring of the variation of the difference in momenta of electron and positron beams 
versus location along the orbit of the CESR storage ring for 5.29~GeV beam energy. 
The sidereal time anisotropy $\Delta c_{1,e}/c_e$ (the combined result from all 24 sets of data)
is below $5.5\times 10^{-15}$ with $\sim$95\% confidence (two-sigma level) as follows from
Eqs.~\ref{eq:amom} and~\ref{eq:amas} with $a_\perp = \sqrt {a_{\perp,x}^2 + a_{\perp,y}^2}$.
This limit is about three times better than previously reported by the most accurate
experiment~\cite{BO2010}.

The accuracy of this experiment is limited by the drift of individual electronics channels and to a lesser extent 
by the readout rate of the associated DAQ.
With resolution of these issues, the technique will allow investigation of the $\Delta c_{1,e}$ 
in the region potentially sensitive to a QG effect.
Performing the experiment at different beam energies will provide an additional handle
for the discrimination of the systematics and the possible AMAS effect.

It is interesting that the LHC collider, whose magnetic systems for two beams are coupled magnetically and
geometrically, provides the possibility of doing a test of AMAS for the proton~\cite{JW2018}.
We would also like to mention that a synchronized measurement of the beam deflection in several storage 
rings with high accuracy opens additional opportunity for a search for the transient effects~\cite{DB2015}, 
which were proposed in a number of models for the physics beyond SM.

Many recent experiments on LIV were interpreted within the framework of the SME theory.
It would be interesting to perform an SME-based analysis of our measurement as well. 
However, in the minimal SME theory there is no sensitivity of the magnetic deflection 
to the dipole form of the AMAS~\cite{PC2015}, 
so the obtained limit could also be used for a test of the underlying assumption of SME.
At the same time, within the SME framework there is already a prediction of a nonvanishing 
eccentricity of the particle trajectory in the magnetic deflection~\cite{BA2011} which could be investigated 
by using the beam trajectory in a storage ring, as we did here, when sufficient stability of the magnets is achieved.
\begin{acknowledgments}
It is our pleasure to thank the CESR staff for the smooth operation of the storage ring.
We would like to extend special thanks to V.~Zelevinsky for his recommendations, 
D.~Sagan and B.~Schmookler for their contributions to data analysis at the initial stage of the experiment, and 
G.~Cates for productive discussions and support of the experiment.
This work supported by National Science Foundation under grants PHYS-1416318, and DGE-1144153.

\end{acknowledgments}

\end{document}